\documentclass[final,1p,times,number]{elsarticle}

\usepackage{amsmath}
\usepackage{amssymb}
\usepackage{amsfonts}
\usepackage{graphicx}
\usepackage{epstopdf}
\usepackage{color}
\usepackage{bm}
\usepackage{multirow}
\usepackage{natbib}
\usepackage{hyperref}

\usepackage{ifpdf}
\ifpdf%
\usepackage{pdflscape}
\else
\usepackage{lscape}
\fi

\DeclareMathOperator{\Tr}{Tr}

\DeclareMathOperator{\spp}{sp}
\DeclareMathOperator{\liq}{li_2}
\DeclareMathOperator{\lif}{li_4}
\DeclareMathOperator{\lik}{li_k}
\DeclareMathOperator{\sgn}{sgn}

\journal{Annals of Physics}

\biboptions{sort&compress}

\begin{document}

\begin{frontmatter}
\title{Two-loop renormalization of the Finkel'stein theory: \\ The specific heat}
\author[ISB1,ISB2]{I.S. Burmistrov\corref{cor}}
\cortext[cor]{Corresponding author. Fax: +7-495-7029317 } 
\ead{burmi@itp.ac.ru}

\address[ISB1]{L.D. Landau Institute for Theoretical Physics,
Kosygina street 2, 119334 Moscow, Russia}
\address[ISB2]{Moscow Institute of Physics and
Technology, 141700 Moscow, Russia}

\begin{abstract}
We explore the two-loop renormalization of the specific heat for an interacting disordered electron system in the case of broken time reversal symmetry. Within the nonlinear sigma model approach we derive the two-loop result for the anomalous dimension which controls scaling of the specific heat with temperature.  As an example, we elaborate the metal-insulator transition in $d=2+\epsilon$ dimensions for the case of broken time reversal and spin rotational symmetries and in the presence of Coulomb interaction. In this situation scaling of the specific heat is determined by the anomalous dimension of the Finkel'stein operator which is the eigen operator of the renormalization group complementary to the eigen operator corresponding  to the second moment of the local density of states. We find that the absolute values of the anomalous dimensions of these operators differ beyond one-loop approximation contrary to the noninteracting case. 
\end{abstract}


\end{frontmatter}

\section{Introduction\label{Sec:Intro}}

The phenomenon of Anderson localization \cite{Anderson1958} has been attracting a lot of interest  for more than 50 years since its discovery (see e.g., \cite{Anderson50years}). The most intricate situation exists in $d=2$ dimension in which, depending on the symmetry class, a noninteracting electron system can be fully localized, fully delocalized or undergoes the Anderson transition. The most convenient tool to study the metallic phase and Anderson transition in noninteracting electron system is the low energy effective action called nonlinear sigma model (NLSM) \cite{Wegner1979,Schaefer1980,Efetov1980,Juengling1980,McKane1981,Efetov1982}. This effective theory 
describes interaction of diffusive modes on scales larger than the mean free path. In $d=2$ this interaction results in logarithmic divergences which are summed by NLSM in a much more convenient way than a standard diagrammatic technique (see e.g., \cite{Efetov1983,Lee1985}). In the presence of both time reversal and spin rotational symmetry (class AI of the Wigner-Dyson classification \cite{Wigner1951,Dyson1962a,Dyson1962b}) the two-dimensional (2D) noninteracting electron system is believed to be always localized at zero temperature: there is no Anderson transition \cite{Anderson1979}.

Inevitably electron-electron interaction becomes important at low temperatures. First of all, inelastic electron-electron scattering with a small (compared to temperature) energy transfer destroys quantum phase coherence  \cite{Thouless1977,Anderson1979b,Altshuler1982}. This leads to a temperature ($T$) dependence of the quantum correction to conductivity \cite{Altshuler1982}.  Additional dependence of conductivity on $T$ appears due to virtual electron-electron scattering \cite{Altshuler1979c}. Physically, this occurs via coherent scattering electrons off the Friedel oscillations \cite{Zala2001}. Strong in comparison with Fermi liquid temperature dependence exists also in such thermodynamic quantities as the specific heat and static spin susceptibility (see e.g., \cite{AAbook}). In $d=2$ both weak localization \cite{Gorkov1979} and electron-electron \cite{Altshuler1980} contributions to conductivity are logarithmic in temperature and opposite in sign. 
This suggests that  the 2D metal-insulator quantum phase transition is possible in the presence of  electron-electron interaction. 

The first extension of the one-parameter scaling theory of Ref. \cite{Anderson1979} to the case of electron-electron interactions was performed in Ref. \cite{McMillan1981}. Although this semi-phenomenological theory suffered from confusion between the local and thermodynamic density of states it put forward an important idea of the two-parameter scaling description of the metal-insulator transition in the presence of electron-electron interaction. Such multi-parameter 
 scaling description was proven to be correct when NLSM has been derived for the case of an interacting electron system \cite{Finkelstein1983a}. With the help of one-loop renormalization group (RG) analysis of this NLSM an interplay of electron-electron interaction and disorder was analyzed \cite{Finkelstein1983b,Finkelstein1984a,Finkelstein1984b,Castellani1984a,Castellani1984b,Finkelstein1984c}. 
In 2D case for the symmetry class AI delocalization due to electron-electron interaction overcomes weak localization in the weak disorder regime. This yields the metallic behavior of conductivity at low temperatures \cite{Finkelstein1984c}. This fact supports existence of 2D metal-insulator transition in the presence of electron-electron interaction.

A change in resistivity from insulating to metallic behavior with increase of electron density was measured in Si metal-oxide-semiconductor field effect transistor \cite{Kravchenko1994,Kravchenko1995}. Similar behavior of resistivity was experimentally observed later in a variety of 2D electron systems (for review, see \cite{Abrahams2001,Dolgopolov2003,Kravchenko2004,Pudalov2004,Shashkin2005}). Observed temperature and electron density dependence of resistivity resembles the expected behavior of resistivity near a 2D metal-insulator transition and was found to be in reasonable agreement with the predictions of the two-parameter scaling theory \cite{JETPL2007,Kravchenko2007,Pudalov2008}. However, recent thermodynamics and transport measurements in Si metal-oxide-semiconductor field effect transistor suggest that the observed strong temperature and electron concentration dependence of resistivity occurs in the regime of nondegenerate Fermi system and, consequently, has nothing to do with the metal-insulator quantum phase transition \cite{Reznikov2010,Teneh2012,Tupikov2015,Kuntsevich2015,Morgun2015}. These new experimental results call for development of the transport theory of nondegenerate strongly interacting 2D electron system, on the one hand, and more detailed understanding of the 2D metal-insulator transition within NLSM approach, on the other hand.

At present, there is not much known on renormalization of the Finkel'stein NLSM beyond one-loop approximation (the lowest order in disorder). There are only few results within two-loop order approximation. The renormalization of the specific heat and static spin susceptibility has been studied 
near the Stoner instability \cite{Kirkpatrick1989,Kirkpatrick1990}. It was demonstrated that there is no metal-insulator transition in 2D electron system with Coulomb interaction and with broken time reversal and spin rotational symmetries \cite{Burmistrov2002}. On the contrary, 
existence of the metal-insulator transition was shown in 2D interacting electron system with $\mathcal{N}\to \infty$ flavors (the action of NLSM is invariant under SU($\mathcal{N}$) rotations) \cite{Finkelstein2005}. 
Recently, existence of multifractality in moments of local density of states in the presence of interactions has been established \cite{Burmistrov2013,Burmistrov2015a}. This situation is in sharp contrast to the knowledge 
on noninteracting NLSM for which beta-function and anomalous dimensions of RG eigen operators are known to the fifth \cite{Hikami1983,Wegner1986a,Wegner1989} and fourth \cite{Wegner1986b,Wegner1987a,Wegner1987b} loop orders, respectively.

In this paper, we  consider the two-loop renormalization of the specific heat for an interacting disordered electron system. For simplicity, we assume that the time reversal symmetry is broken, e.g. by a weak magnetic field. This assumption allows us to avoid contributions from the Cooper channel. In the absence of electron-electron interaction the system under consideration belongs to the symmetry class A. We derive the two-loop RG equation for the Finkel'stein parameter which determines the temperature behavior of the specific heat. In the case of additionally broken spin rotational symmetry we compare the anomalous dimension of the Finkel'stein operator in the NLSM action with the anomalous dimension of the second moment of the local density of states. We find that (i) these anomalous dimensions have opposite sign and (ii) the absolute values of these anomalous dimensions are different beyond one-loop approximation.

The paper is organized as follows. In Sec. \ref{Sec:Form} we introduce NLSM approach. Next we present details of the two-loop computations of the Finkel'stein parameter in $d=2+\epsilon$ dimensions (Sec. \ref{Sec:2loop}). In Sec. \ref{Sec:SA} we consider the metal-insulator transition in $d=2+\epsilon$ dimensions in the electron system with broken time reversal and spin rotational symmetries. We end the paper with Conclusions (Sec. \ref{Sec:Conc}). Some additional details of two-loop calculations are given in \ref{App1}.

\section{Formalism\label{Sec:Form}}

\subsection{Nonlinear sigma model action}

For the case of preserved spin rotational but broken time reversal symmetries the action of NLSM is given as the sum of the noninteracting part, $S_\sigma$,
and contributions arising from the interactions in the particle-hole singlet and triplet channels, $S_{\rm int}$ (for review, see \cite{Finkelstein1990,BelitzKirkpatrick1994}):
\begin{gather}
S=S_\sigma + S_{\rm F} .
\label{eq:NLSM}
\end{gather}
Here the noninteracting part is given as
\begin{equation}
S_\sigma  = -\frac{g}{16} \int d\bm{r} \Tr (\nabla Q)^2 ,
\label{Ss}
\end{equation}
where $g=2\pi \nu D$ is the total Drude conductivity (in units $e^2/h$ and including spin).  The Finkel'stein part of the action which involves interaction is as follows   
\begin{gather}
S_{\rm F}  =- \frac{\pi T}{2} \sum_{\alpha,n} \sum_{j=0}^3\Gamma_j 
\int d\bm{r} \Tr \Bigl [I_n^\alpha \sigma_{j} Q\Bigr ]  \Tr \Bigl [I_{-n}^\alpha \sigma_{j} Q\Bigr ] 
+ 4\pi T z_\omega \int d\bm{r} \Tr \eta ( Q - \Lambda)\notag \\
 - 2\pi T z_\omega \int d\bm{r} \Tr \eta  \Lambda .
\label{Srho}
\end{gather}
Here $\Gamma_0 = \Gamma_s$ and $\Gamma_1=\Gamma_2=\Gamma_3=\Gamma_t$ denote the interaction amplitudes in the particle-hole singlet and triplet channels, respectively. The parameter $z_\omega$ is frequency renormalization factor introduced by Finkel'stein \cite{Finkelstein1983a}. 
 We use the following matrices
\begin{gather}
\Lambda_{nm}^{\alpha\beta} = \sgn n \, \delta_{nm} \delta^{\alpha\beta}\sigma_{0}, \quad
\eta_{nm}^{\alpha\beta}=n \, \delta_{nm}\delta^{\alpha\beta} \sigma_{0},  \quad
(I_k^\gamma)_{nm}^{\alpha\beta}=\delta_{n-m,k}\delta^{\alpha\beta}\delta^{\alpha\gamma} \sigma_{0} , 
\end{gather}
where $\alpha,\beta = 1,\dots, N_r$ are replica indices. Integer numbers $n$ and $m$ correspond to the
Matsubara fermionic energies $\varepsilon_n = \pi T (2n+1)$ and $\varepsilon_m = \pi T (2m+1)$. The four Pauli matrices, 
\begin{equation}
\sigma_0 = \begin{pmatrix}
1 & 0\\
0 & 1
\end{pmatrix},\quad
\sigma_1  = \begin{pmatrix}
0 & 1\\
1 & 0
\end{pmatrix}, \quad \sigma_2 = \begin{pmatrix}
0 & -i\\
i & 0
\end{pmatrix}, \quad \sigma_3 = \begin{pmatrix}
1 & 0\\
0 & -1
\end{pmatrix} , \label{eq:tau-def}
\end{equation}
operate in the spin space. The matrix field $Q(\bm{r})$ acting in the replica, Matsubara,
and spin spaces obeys the following constraints: $Q^2=1$ and $\Tr Q = 0$.

\subsection{$\mathcal{F}$-algebra and $\mathcal{F}$-invariance}

The NLSM action \eqref{eq:NLSM} involves the matrices in the Matsubara frequency space. Formally, Matsubara frequencies runs from minus to plus infinity which makes matrices of infinite size. To perform actual calculations with such matrices  we introduce an ultraviolet cutoff  $N_M^\prime$ for the Matsubara frequencies. Following Ref. \cite{PruiskenBaranovSkoric1999}, we introduce additional cutoff $N_M < N_M^\prime$ which separates non-trivial and trivial (beyond which the $Q$ matrix equals $\Lambda$) parts of the $Q$ matrix. At the end of calculations the limit $N_M, N_M^\prime \to \infty$ should be taken.

As known \cite{PruiskenBaranovSkoric1999,Kamenev1999}, rotations of the $Q$ matrix with a matrix $\exp (i \hat \chi)$ where $\hat\chi = \sum_{\alpha,n} \chi_n^\alpha I^\alpha_n \sigma_{0}$ play an important role. Such rotations correspond to the gauge transformations in the original fermionic language. 
In the limit $N_M, N_M^\prime \to \infty$ and $N_M/N_M^\prime \to 0$,
the set of rules known as $\mathcal{F}$ algebra \cite{PruiskenBaranovSkoric1999}
allows one to establish the following relations ($j=0,1,2,3$):
\begin{equation}
\begin{split}
\Tr I^\alpha_n \sigma_{j}e^{i \hat\chi} Q e^{-i \hat\chi} & = \Tr I^\alpha_n \sigma_{j} Q  + 4 i n \chi_{-n}^\alpha \delta_{j0},  \\
\Tr \eta e^{i \hat\chi} Q e^{-i \hat\chi} & = \Tr \eta Q  + \sum_{\alpha, n} i n \chi_{n}^\alpha
\Tr I^\alpha_n \sigma_{0} Q - 4  \sum_{\alpha, n} n^2 \chi_{n}^\alpha
\chi_{-n}^\alpha .
\end{split}
\label{eq:Falg}
\end{equation}
Using Eq. \eqref{eq:Falg}, one can check that, provided $\Gamma_s=-z_\omega$, the NLSM action is invariant under global rotations of the matrix $Q$ with the matrix $\exp (i \hat \chi)$ (so called $\mathcal{F}$ invariance). We remind that the constraint $\Gamma_s=-z_\omega$ corresponds to the case of Coulomb interaction \cite{Finkelstein1983a}. Since the relation $\Gamma_s=-z_\omega$ allows additional symmetry of the NLSM action, this relation remains fulfilled under the RG flow.

\subsection{Thermodynamic potential}

The thermodynamic potential per unit volume is determined by the NLSM action:
\begin{equation}
\Omega = - \frac{T}{V} \ln \int \mathcal{D}[Q] \exp S \, ,
\end{equation}
where $V$ stands for a sample volume. At the classical level, for 
$Q=\Lambda$,  the thermodynamic potential is equal to $\Omega = -T S_F[\Lambda] = 2\pi z_\omega T^2 \Tr \eta \Lambda$. The quantum corrections to the thermodynamic potential determine the renormalized value of the frequency renormalization parameter \cite{Baranov1999}:
\begin{equation}
z^\prime_\omega = \frac{1}{2\pi \Tr \eta \Lambda} \frac{\partial}{\partial T} \frac{\Omega}{T} = 
z_\omega \frac{\langle S_F[Q] \rangle}{S_F[\Lambda]} .
\label{eq:Zp:ren}
\end{equation}
We note that $z^\prime_\omega$ is responsible for the non Fermi-liquid temperature behavior of the specific heat \cite{Castellani1986}.

\section{Two-loop renormalization of $z_\omega$ \label{Sec:2loop}}

\subsection{Perturbative expansion}

For the perturbative treatment (in $1/g$) of the NLSM action \eqref{eq:NLSM}
we shall use the square-root parametrization
\begin{gather}
Q = W +\Lambda \sqrt{1-W^2}\ , \qquad W= \begin{pmatrix}
0 & w\\
\bar{w} & 0
\end{pmatrix} .
\label{eq:Q-W}
\end{gather}
The blocks $w$ and $\bar{w}$ are independent matrix variables. 
They have the following nonzero elements in the Matsubara space: $w_{n_1n_2}$ and $\bar{w}_{n_2n_1}$ with $n_1\geqslant 0$ and $n_2< 0$. It is convenient to represent $w$ and $\bar{w}$ as the linear combinations of the Pauli matrices: $w^{\alpha\beta}_{n_1n_2}= \sum_{j} (w^{\alpha\beta}_{n_1n_2})_{j} \sigma_{j}$ and $\bar{w}^{\beta\alpha}_{n_2n_1}= \sum_{j} (w^{\beta\alpha}_{n_2n_1})_{j} \sigma_{j}$.
In what follows, we use the convention: $n_1, n_3, n_5, \dots \geqslant 0$ and $n_2, n_4, n_6, \dots <0$.

Expanding the NLSM action \eqref{eq:NLSM} to the second order in $W$, we obtain the following propagators for diffusive modes ($j=0,1,2,3$):
\begin{gather}
\Bigl \langle [w_{j}(\bm{q})]^{\alpha_1\beta_1}_{n_1n_2} [\bar{w}_{j}(-\bm{q})]^{\beta_2\alpha_2}_{n_4n_3} \Bigr \rangle =  \frac{4}{g} \delta^{\alpha_1\alpha_2} \delta^{\beta_1\beta_2}\delta_{n_{12},n_{34}}
\mathcal{D}_q(i\omega_{12})\Biggl [\delta_{n_1n_3} - \frac{16 \pi T \Gamma_j}{g}\delta^{\alpha_1\beta_1}  \mathcal{D}_q^{(j)}(i\omega_{12}) \Biggr ] ,
\label{eq:prop:PH}
\end{gather}
where $\bm{q}$ stands for the momentum, {$n_{12}= n_1-n_2$} and $\omega_{12} = \varepsilon_{n_1}-\varepsilon_{n_2}$. The standard propagator for diffuson is given as ($\omega_n = 2\pi T n$):
\begin{equation}
\mathcal{D}^{-1}_q(i\omega_n) =q^2+{8 z_\omega |\omega_n|}/{g} .
\label{eq:prop:Free}
\end{equation}
The diffusive modes renormalized by interaction in the singlet  ($\mathcal{D}_q^{(0)}(i\omega) \equiv \mathcal{D}_q^{s}(i\omega)$) and triplet  ($\mathcal{D}_q^{(1)}(i\omega)=\mathcal{D}_q^{(2)}(i\omega)=\mathcal{D}_q^{(3)}(i\omega) \equiv \mathcal{D}_q^{t}(i\omega)$)   particle-hole channels are as follows
\begin{equation}
\begin{split}
[\mathcal{D}^s_q(i\omega_n)]^{-1} & =  q^2+{8 (z_\omega+\Gamma_s) |\omega_n|}/{g},\\
 [\mathcal{D}^t_q(i\omega_n)]^{-1} & =  q^2+{8 (z_\omega+\Gamma_t) |\omega_n|}/{g} .
 \end{split}
 \label{eq:prop:Int}
\end{equation}

For the purpose of regularization in the infrared, it is convenient to add the 
term to the NLSM action:
\begin{equation}
S \to S + \frac{g h^2}{8} \int d \bm{r}\Tr \Lambda Q .
\label{SsGenFull}
\end{equation}
This leads to the shift of the momentum squared,  $q^2 \to q^2 +h^2$, in the propagators \eqref{eq:prop:Free} and \eqref{eq:prop:Int}.

\subsection{One-loop perturbative results}

Before going to the two-loop results we remind briefly the one-loop perturbative results for $z_\omega$. 
Expanding the NLSM action to the second order in $W$ and using Eq. \eqref{eq:prop:PH}, we find the one-loop perturbative result:
\begin{equation}
\frac{\partial}{\partial T} \frac{\Omega^{(1)}}{T} = - \frac{S_F[\Lambda]}{TV} \left ( 1 + \frac{2}{g} \sum_{j=0}^3 \gamma_j \int_q \mathcal{D}_q^{(j)}(0)\right ) .
\end{equation}
Here we use the following notations: $\gamma_j = \Gamma_j/z_\omega$ and $
\int_q \equiv \int d^d\bm{q}/(2\pi)^d$. Evaluating integral over momentum in $d=2+\epsilon$ dimensions 
and using Eq. \eqref{eq:Zp:ren}, we obtain \cite{Finkelstein1983a}:
\begin{equation}
z_\omega^\prime = z_\omega \left ( 1 - \frac{t h^\epsilon}{\epsilon} (\gamma_s+3\gamma_t) \right ) .
\label{eq:Z:1l}
\end{equation}
Here we introduce resistivity $t = 4\Omega_d/g$, where $\Omega_d= S_d/[2(2\pi)^d]$ and $S_d= 2\pi^{d/2}/\Gamma(d/2)$ is the area of the $d$-dimensional sphere.

\subsection{Two-loop perturbative results}

The two-loop contribution to the thermodynamic potential can be written as 
\begin{equation}
\Omega^{(2)} = -\frac{T}{V} \left \langle S^{(4)}_0 + S^{(4)}_{\rm int} + \frac{1}{2} \left (S^{(3)}_{\rm int} \right )^2 \right \rangle .
\end{equation}
Here the term, 
\begin{equation}
S^{(3)}_{\rm int}   = \frac{\pi T}{2} \sum_{j=0}^3 \Gamma_j \sum_{\alpha,n} \int d\bm{r} \Tr I^{\alpha}_{n} \sigma_{j} W \Tr I^{\alpha}_{-n} \sigma_{j}\Lambda W^2,
\end{equation}
appears from the expansion of the NLSM action to the third order in $W$. The forth order terms are given as
\begin{align}
S^{(4)}_0   = & -\frac{g}{64} \int_{q_j} \delta\left (\sum_{j=0}^3\bm{q_j}\right ) 
\sum_{\beta_1\beta_2\beta_3\beta_4}\sum_{n_5n_6n_7n_8}   \spp \Bigl [ w^{\beta_1\beta_2}_{n_5n_6}(\bm{q_0}) \bar{w}^{\beta_2\beta_3}_{n_6n_7}(\bm{q_1})
w^{\beta_3\beta_4}_{n_7n_8}(\bm{q_2}) \bar{w}^{\beta_4\beta_1}_{n_8n_5}(\bm{q_3})\Bigl ]\notag \\
& \times
 \Bigl [ 2h^2+ \frac{16 z_\omega}{g} (\omega_{56}+\omega_{78})-(\bm{q_0}+\bm{q_1})(\bm{q_2}+\bm{q_3})
 -(\bm{q_0}+\bm{q_3})(\bm{q_1}+\bm{q_2}) \Bigr ]  
\end{align}
and 
\begin{equation}
S^{(4)}_{\rm int}  = -\frac{\pi T}{8}\sum_{j=0}^3 \Gamma_j \sum_{\alpha,n} \int d\bm{r} \Tr I^{\alpha}_{n} \sigma_{j} \Lambda W^2  
\Tr I^{\alpha}_{-n} \sigma_{j}\Lambda W^2 .
\end{equation} 
The symbol $\spp$ denotes the trace over the spin space only. Performing contraction with the help of the Wick theorem and Eq. \eqref{eq:prop:PH}, we obtain
\begin{gather}
\langle S^{(4)}_0 \rangle = - \frac{2 \pi T N_r V}{g}\left (\frac{8}{g}\right )^2 \int_{p,q} \sum_{m,n>0} \min\{\omega_m,\omega_n\}
\Bigl [\mathcal{D}^{-1}_q(i\omega_n)+\mathcal{D}^{-1}_p(i\omega_m) \Bigr ]
\sum_{j=0}^3 \Gamma_j  \mathcal{D}_q(i\omega_n) \mathcal{D}^{(j)}_q(i\omega_n) \notag \\
\times 
\sum_{j^\prime=0}^3 \Gamma_{j^\prime} \mathcal{D}_p(i\omega_m) \mathcal{D}^{(j^\prime)}_p(i\omega_m) ,
\label{eq:2l:1}
\end{gather}
\begin{align}
\langle S^{(4)}_{\rm int} \rangle =  & - \frac{4 \pi T N_r V}{g}\left (\frac{8}{g}\right )^2 \int_{p,q} \sum_{m,n>0} \min\{\omega_m,\omega_n\} \mathcal{D}_q(i\omega_{m+n})
\Biggl [ 3 \Gamma_t (\Gamma_s-\Gamma_t) \mathcal{D}_p(i\omega_m) \mathcal{D}^{t}_p(i\omega_m)  
\notag \\
& + \Gamma_s(\Gamma_s+3\Gamma_t) \mathcal{D}_p(i\omega_m) \mathcal{D}^{s}_p(i\omega_m) \Biggr ] 
+ \frac{32 \pi T N_r V}{g^2} \Gamma_s \int_{p,q} \sum_{m,n>0} \omega_m  \mathcal{D}_p(i\omega_m)\mathcal{D}_q(i\omega_m) 
\notag \\
& + \frac{48 \pi T N_r V}{g}\left (\frac{8}{g}\right )^2 \int_{p,q} \sum_{m,n>0} \omega_m 
\Biggl [ \mathcal{D}_p(i\omega_m)+ \mathcal{D}^t_q(i\omega_{m+n})\Biggr ]\mathcal{D}^{t}_p(i\omega_m) \mathcal{D}_q(i\omega_{m+n}) 
,\label{eq:2l:2}
\end{align}
and
\begin{align}
\left \langle \frac{1}{2}\left ( S^{(3)}_{\rm int} \right )^2 \right \rangle =  &
\frac{16\pi T N_r V}{g} \left (\frac{8}{g}\right )^2 \int_{p,q} \sum_{m,n>0} \min\{\omega_m,\omega_n\} 
\mathcal{D}_{\bm{p+q}}(i\omega_{m+n})
\Biggl \{ 
\frac{2 \omega_n}{g} \Bigl [ \Gamma_t^2 \mathcal{D}^t_q(i\omega_{n})
-\Gamma_s^2 \mathcal{D}^s_q(i\omega_{n}) \Bigr ] \notag \\
& \times \sum_{j=0}^3 \Gamma_{j} \mathcal{D}_p(i\omega_m) \mathcal{D}^{(j)}_p(i\omega_m) 
 - \frac{8 \omega_n}{g} \Gamma_s \Gamma_t^2 \mathcal{D}^t_q(i\omega_{n}) \mathcal{D}_p(i\omega_m) \mathcal{D}^s_p(i\omega_m) \notag \\
& + \frac{1}{4}\Bigl [\Gamma_s^2 \mathcal{D}^s_q(i\omega_{n}) \mathcal{D}^s_p(i\omega_m)+
6 \Gamma_s\Gamma_t \mathcal{D}^s_q(i\omega_{n}) \mathcal{D}^t_p(i\omega_m)  -
3 \Gamma_t^2\mathcal{D}^t_q(i\omega_{n}) \mathcal{D}^t_p(i\omega_m)\Bigr ] 
\Biggr \} 
\notag \\
& - \frac{48\pi T N_r V}{g} \left (\frac{8}{g}\right )^3 \int_{p,q} \sum_{m,n>0} \omega_m \omega_n
\Gamma_t^3 \mathcal{D}^t_q(i\omega_{n}) \mathcal{D}_p(i\omega_m)\mathcal{D}^t_p(i\omega_m) \mathcal{D}_{\bm{p+q}} .
\label{eq:2l:3}
\end{align}

Next we combine the two-loop contributions \eqref{eq:2l:1} -- \eqref{eq:2l:3} and rewrite them in the form $\sum_{m>0} \omega_m Y(i\omega_m)$ with the help of following identity: 
$\sum_{m,n>0}\min\{m,n\} y(m,n) = \sum_{m,n>0} m \bigl [y(m,n+m)+y(n+m,m)\bigr ]$. Taking the limit $\omega_m \to 0$ and evaluating the momentum integrals and Matsubara frequency sum, we find (see \ref{App1})
\begin{align}
z_\omega^\prime = z_\omega \Biggl ( 1+\frac{b_1 h^{\epsilon}t}{\epsilon}
+\frac{h^{2\epsilon}t^2}{\epsilon^2} \bigl (b_2+\epsilon {b}_3\bigr )  + O(t^3)\Biggr ) .
\label{eq:Z:pert:2l}
\end{align}
Here we add the one-loop contribution \eqref{eq:Z:1l} and introduce the following functions of interaction parameters:
\begin{equation}
b_1 = - \gamma_s - 3 \gamma_t \, ,
\qquad b_2 = (3\gamma_t+\gamma_s) \bigl [ f(\gamma_s)+3f(\gamma_t)\bigr ] 
+6\gamma_t^2 - 2\gamma_s \, ,
\end{equation}
and
\begin{gather}
{b}_3 = \frac{3\gamma_t+\gamma_s}{2} \sum_{j=0}^3 \Biggl
[2 f(\gamma_j)+\ln(1+\gamma_j) -2- \frac{(1+\gamma_j)\ln^2(1+\gamma_j)}{2\gamma_j}
-\frac{2+3\gamma_j}{\gamma_j}\liq(-\gamma_j) \Biggr ]
\notag \\
-
6\gamma_t[\gamma_t-\liq(-\gamma_t)] \, .
\end{gather}
Here $f(x) = 1 - (1+1/x)\ln(1+x)$ and $\lik(x) = \sum_{m=1}^\infty x^m/m^k$ stands for the polylogarithm (Jonqui\`ere's function).

\subsection{Anomalous dimension of $z_\omega$}

Since operator $\Tr \Lambda Q$ determines the local single-particle density of states, the momentum scale $h$ acquires renormalization \cite{Baranov1999}. The corresponding renormalized momentum scale $h^\prime$ is defined according to
\begin{equation}
g h^2 \langle \Tr \Lambda Q \rangle = g^\prime h^{\prime 2} \Tr \Lambda^2 ,
\end{equation}
where $g^\prime$ stands for the renormalized conductivity. Within the one-loop approximation, one can find \cite{Baranov1999}
\begin{equation}
h^\prime = h \Biggl [ 1 - \frac{h^\epsilon t}{2\epsilon} \sum_{j=0}^3\Bigl ( 2 f(\gamma_j)+\ln(1+\gamma_j)\Bigr )\Biggr ] ,
\label{eq:h:1l}
\end{equation}
and \cite{Altshuler1979c,Altshuler1980,AAbook,Finkelstein1983a,Castellani1984a}
\begin{equation}
g^\prime = g \Biggl [ 1+ a_1 \frac{h^\epsilon t}{\epsilon} \Biggr ] , \qquad 
a_1 = 2 \sum_{j=0}^3 f(\gamma_j) .
\label{eq:g:1l}
\end{equation}
Also we remind the one-loop results for the renormalization of interaction parameters \cite{Finkelstein1983a,Castellani1984a}:
\begin{equation}
\begin{split}
\gamma_s^\prime & = \gamma_s +  c_{s,1} \frac{h^\epsilon t}{\epsilon}, \qquad c_{s,1}=(1+\gamma_s)(3\gamma_t+\gamma_s), \\ 
\gamma_t^\prime & = \gamma_t + c_{t,1} \frac{h^\epsilon t}{\epsilon}, \qquad  c_{t,1} =(1+\gamma_t)(\gamma_s-\gamma_t) .
\end{split}
\label{eq:gg:1l}
\end{equation}

In order to extract the anomalous dimension of $z_\omega$ from the perturbative result \eqref{eq:Z:pert:2l}, 
we use the minimal subtraction scheme \cite{Amit-book}. Let us introduce the dimensionless resistance $\bar{t} = t^\prime h^{\prime \epsilon}$. Then using Eqs. \eqref{eq:h:1l}, \eqref{eq:g:1l}, \eqref{eq:gg:1l} and \eqref{eq:Z:pert:2l},
we express $t$, $\gamma_j$ and $z_\omega$ via $\bar{t}$, $\gamma_j^\prime$ and $z_\omega^\prime$:
\begin{equation}
t = (h^\prime)^{-\epsilon} \bar{t} \, {Z}_t(\bar{t}, \gamma_s^\prime,\gamma_t^\prime), 
\qquad \gamma_j = \gamma_j^\prime \, {Z}_{\gamma_j}(\bar{t}, \gamma_s^\prime,\gamma_t^\prime), \qquad
z_\omega = z_\omega^\prime \, {Z}_{z_\omega}(\bar{t}, \gamma_s^\prime,\gamma_t^\prime) .
\end{equation}
To the first order in $\bar{t}$ we find
\begin{equation}
Z_t = 1+a_1 \frac{\bar{t}}{\epsilon}, \qquad 
Z_{\gamma_j} = 1 - \frac{c_{j,1}}{\gamma_j^\prime} \frac{\bar{t}}{\epsilon} ,
\end{equation}
where $a_1$, $c_{s,1}$, and $c_{t,1}$ are the functions of $\gamma_s^\prime, \gamma_t^\prime$ now. To the second order in $\bar{t}$ the function ${Z}_{z_\omega}$ becomes
\begin{equation}
Z^{-1}_{z_\omega} = 1+ \frac{b_1 \bar{t}}{\epsilon}
+\frac{\bar{t}^2}{\epsilon^2} \left (b_2 + a_1 b_1 - \sum_{j=0}^3\frac{\partial b_1}{\partial \gamma_j^\prime} c_{j,1} + \epsilon \Bigl ( {b}_3 + b_1 \sum_{j=0}^3 \bigl [ f(\gamma_j) +\frac{1}{2}\ln(1+\gamma_j)\bigr ] \Bigr )\right ) .
\end{equation}
Here $a_1$, $b_{1,2,3}$, $c_{s,1}$, and $c_{t,1}$ are the functions of $\gamma_s^\prime, \gamma_t^\prime$. The RG equations can be found from the standard condition that $t$, $\gamma_{s,t}$ and $z_\omega$ are independent of $h^\prime$. In this way we obtain the following two-loop result for the anomalous dimension of $z_\omega$:
\begin{equation}
- \frac{d \ln z_\omega}{d y} =\zeta_z(t,\gamma_s,\gamma_t) = - t (\gamma_s+3\gamma_t) 
- t^2 \Biggl [ (\gamma_s+3\gamma_t) \sum_{j=0}^3 \Bigl (c(\gamma_j) + 2\liq(-\gamma_j)\Bigr ) + 12 \gamma_t\Bigl (\gamma_t-\liq(-\gamma_t)\Bigr ) \Biggr ] ,
\label{eq:ad:2l}
\end{equation} 
where $y = \ln h/h^\prime$ and we introduce the function (see Refs. \cite{Burmistrov2013,Burmistrov2015a})
\begin{equation}
c(\gamma) = 2 + \frac{2+\gamma}{\gamma}\liq(-\gamma)+\frac{1+\gamma}{2\gamma^2}\ln^2(1+\gamma) .
\label{eq:c:def}
\end{equation}
For a brevity, we omitted the prime and bar signs. Equation \eqref{eq:ad:2l} is the main result of our paper. We note that the relation 
\begin{equation}
b_2 = \frac{1}{2} \left ( b_1^2 -a_1 b_1 + \sum_{j=0}^3\frac{\partial b_1}{\partial \gamma_j^\prime} c_{j,1} \right )
\end{equation}
holds. This guarantees the renormalizability of the theory within the two-loop approximation, i.e. the absence of terms in the right hand side of Eq. \eqref{eq:ad:2l} which diverge in the limit $\epsilon \to 0$.

We note that in the limit $\gamma_t\gg 1$ (Stoner instability corresponds to $\gamma_t=\infty$), the anomalous dimension \eqref{eq:ad:2l} becomes $\zeta = -3 \gamma_t t -12 \gamma_t^2 t^2$ in agreement with the result of Belitz and Kirkpatrick (see Eq. (6.60c) from Ref. \cite{BelitzKirkpatrick1994}). This result indicates that toward Stoner instability the loop expansion is controlled by the small parameter $\gamma_t t \ll 1$ rather than $t\ll 1$ as in the case of noninteracting electrons. The two-loop result \eqref{eq:ad:2l} for the anomalous dimension of $z_\omega$ interpolates between the result of Ref. \cite{Baranov1999} for the case when the interaction in the triplet channel is absent, $\gamma_t=0$ and the result of Refs. \cite{Kirkpatrick1989,Kirkpatrick1990} for the case $\gamma_t \to \infty$.

The following remark is in order here. As a consequence of the particle number conservation, the quantity $z+\Gamma_s$ has no renormalization \cite{Finkelstein1983a}. Therefore, the renormalization of the interaction parameter $\gamma_s$ is fully determined by the anomalous dimension $\zeta_z$:
\begin{equation}
- \frac{d \gamma_s}{d y} = \beta_{\gamma_s} = - (1+\gamma_s) \zeta_z(t,\gamma_s,\gamma_t) .
\end{equation}
Thus we also derived the two-loop RG equation for the singlet channel interaction parameter $\gamma_s$. 

\section{Scaling analysis \label{Sec:SA}}

As example of application of our result \eqref{eq:ad:2l}, we consider the case of Coulomb interaction, $\gamma_s=-1$ and fully broken spin rotational symmetry such that the triplet channel is absence. Since еру time reversal symmetry is also absent, this situation can be realized in the system of disordered interacting fermions with magnetic impurities. In notations of Ref. \cite{BelitzKirkpatrick1994}, this case is referred as the symmetry class ``MI(LR)''. The anomalous dimension \eqref{eq:ad:2l} becomes
\begin{equation}
\zeta_z(t) =  t + \left (2+\frac{\pi^2}{6}\right ) t^2 + O(t^3).
\label{eq:ad:2l:milr}
\end{equation}
We mention that the numerical coefficient $(2+\pi^2/6)$ in front of the $t^2$ term is different from the result $(3+\pi^2/6)$ found in Ref. \cite{Baranov1999}. We suppose that this mismatch is due to the erroneous treatment of terms singular in $1/(1+\gamma_s)$ in Ref. \cite{Baranov1999} (see \ref{App1}).

The RG equation for the dimensionless resistance $t$ is known up to the two-loop order \cite{Burmistrov2002}:
\begin{equation}
-\frac{dt}{d\ln y}  = \beta_t = \epsilon t - 2 t^2 - 4 A t^3 + O(t^4) ,
\label{Scal_eq1}
\end{equation}
where the constant 
\begin{gather}
A =\frac{1}{16}\Biggl [\frac{139}{6}+\frac{(\pi^2-18)^2}{12}+\frac{19}{2}\zeta (3)+\Bigl ( 16 + \frac{\pi ^2}{3} \Bigr )\ln ^{2}2  - \Bigl (44-\frac{\pi ^{2}}{2}+7\zeta (3)\Bigr ) \ln 2 \notag \\
+16\mathcal{G}-\frac{1}{3}\ln ^{4}2-8\lif \left(\frac{1}{2}\right)\Biggr ] \approx 1.64 .
\end{gather}
Here $\mathcal{G} \approx 0.915$ stands for the Catalan constant and $\zeta(3)\approx 1.2$ denotes the Riemann zeta. As usual, the zero of the $\beta$-function, $\beta_t(t_*)=0$, determines the critical point in $d=2+\epsilon$ dimensions:
\begin{equation}
t_* = \frac{\epsilon}{2} (1-A\epsilon) +O(\epsilon^3) .
\end{equation}
This critical point separates the metallic ($t<t_*$) and insulating ($t>t_*$) phases. At this critical point the 
correlation/localization length diverges
 \begin{equation}
\xi = h^{-1} |t-t_*|^{-\nu},\qquad  \nu = -\left [ \frac{d \beta_t}{dt} \Biggl |_{t=t_*}\right ]^{-1} = \frac{1}{\epsilon} - A + O(\epsilon) .
 \end{equation}
The value of the anomalous dimension 
\begin{equation}
\zeta_z^*=\zeta_z(t_*) =\frac{\epsilon}{2} + \frac{\epsilon^2}{4} \left ( 2-2A+\pi^2/6\right ) + O(\epsilon^3)
\end{equation}
 at the critical point determines the dynamical exponent 
\begin{equation}
z= d-\zeta_z^* = 2+\frac{\epsilon}{2} + \frac{\epsilon^2}{4} \left ( 2A-2-\pi^2/6\right ) + O(\epsilon^3) .
\label{eq:z:2l}
\end{equation}
We remind that the dynamical exponent $z$ determines the length scales induced by energy and temperature, 
$L_E \sim |E|^{-1/z}$ and $L_T \sim T^{-1/z}$, at the critical point. Interestingly, the very same dynamical exponent is responsible for the deviation of the specific heat $c_v$ from the Fermi-liquid-type behavior, $c_v \sim T^{d/z}$ \cite{Castellani1986}. 

In the considered case of the symmetry class MI(LR) the anomalous dimension of the Finkel'stein parameter $\zeta_z$ determines the renormalization of the Finkel'stein part of the NLSM action. The term $S_F$ is an example of the operator bilinear in the matrix $Q$. This operator is $\mathcal{F}$ invariant, local and does not involve spatial gradients, consequently, it is also invariant with respect to spatial rotations of the matrix $Q$ by matrix $\exp(i \hat \chi)$. Such spatial and time dependent rotations of $Q$ correspond to the gauge transformation of the original fermions \cite{PruiskenBaranovSkoric1999}. Since the anomalous dimension of $z_\omega$ is finite in the limit $\epsilon \to 0$ within the two-loop approximation, the operator $S_F$ is eigen operator with respect to the RG. We emphasize that $\zeta_z(t) >0$ in the two-loop approximation.

Recently, it was shown \cite{Burmistrov2013,Burmistrov2015a} that the polinomial in $Q$ operators corresponding to the moments of the local density of states are eigen operators of the RG. The second moment of the local density of states is expressed in terms of the operator $K_2$ which is bilinear in $Q$ similar to the Finkel'stein term. However, contrary to  $S_F$, the operator $K_2$ is not $\mathcal{F}$ invariant due to lack of gauge invariance in the local density of state. For the symmetry class MI(LR) the anomalous dimension of the operator $K_2$ within two-loop approximation is given as \cite{Burmistrov2013}:
\begin{equation}
\zeta_2(t) =  -t - \left (2-\frac{\pi^2}{6}\right ) t^2 + O(t^3).
\label{eq:ad2:2l:milr}
\end{equation}
At the critical point the anomalous dimension of $K_2$ is negative:
\begin{equation}
\Delta_2 = \zeta_2(t_*) =
-\frac{\epsilon}{2} - \left (2-2A-\frac{\pi^2}{6}\right ) \frac{\epsilon^2}{4} +O(\epsilon^3) .
\label{eq:d2}
\end{equation}
In the two-loop approximation the anomalous dimension of the $q$-th moment of the local density of states is 
expressed via the anomalous dimension of the second moment, $\Delta_q = [q(q-1)/2] \Delta_2$. They are  negative and nonlinear functions of $q$. Thus the moments of the local density of states demonstrate the multifractal behavior in the presence of Coulomb interaction.  We mention that within one-loop approximation the anomalous dimensions $\zeta_2$ and $\zeta_z$ are the same except the sign. At the two-loop order they become essentially different.   

One can construct the polynomial in $Q$ eigen operators which corresponds to the higher moments of the local density of states \cite{Burmistrov2013,Burmistrov2015a}. Similarly, one can study
the higher order in $Q$ eigen operators which are $\mathcal{F}$ invariant. However,  at present they are not known beyond the operators with four $Q$ matrices \cite{Pruisken}.   

It is instructive to compare the scaling results obtained above for the case of Coulomb interaction with 
the results for the system of fermions with short-range singlet interaction ($-1<\gamma_s\leqslant 0$). 
The case of short-ranged singlet interaction lies in an attraction region of the noninteracting fixed point. 
In the absence of interaction the symmetry class MI(LR) is just the unitary Wigner-Dyson class A. 
The Anderson transition in the class A is described by the following $\beta$-function
\cite{Hikami1983,Wegner1986a,Wegner1989}
\begin{equation}
\beta_t^{(0)} = \epsilon t - \frac{1}{2} t^3 - \frac{3}{8} t^5 + O(t^6)  .
\end{equation}
The critical point and correlation length exponent are given as
\begin{equation}
t_*^{(0)} = (2\epsilon)^{1/2}\left (1-\frac{3\epsilon}{4}\right )+O(\epsilon^{5/2}), \qquad \nu^{(0)} = \frac{1}{2\epsilon} - 
\frac{3}{4} +O(\epsilon) .
\end{equation}
The anomalous dimensions of the operator $K_2$ within the four-loop approximation is as follows \cite{Wegner1986b,Wegner1987a,Wegner1987b}
\begin{equation}
\zeta_2^{(0)}(t) = - t -  \frac{3 t^3}{8} - \frac{3 \zeta(3)}{8} t^4 +  O(t^5) ,
\end{equation}
In the noninteracting case the multifractal exponent for the operator $K_2$ becomes 
\begin{equation}
\Delta_2^{(0)} = \zeta_2^{(0)}(t_*^{(0)}) =- \left (2 \epsilon \right )^{1/2} - \frac{3 \zeta(3)}{2} \epsilon^2  +  O(\epsilon^{5/2}).
\end{equation}
In addition, to the operator $K_2$ there is the other eigen operator bilinear in $Q$. Its anomalous dimension is also known up to the four-loop order \cite{Wegner1986b,Wegner1987a,Wegner1987b}:
\begin{equation}
\mu_2^{(0)}(t) = t +  \frac{3 t^3}{8} - \frac{3 \zeta(3)}{8} t^4 +  O(t^5) ,
\end{equation}
The corresponding critical exponent is given as
\begin{equation}
\mu_2^{*} = \mu_2^{(0)}(t_*) = \left (2 \epsilon\right )^{1/2}  - \frac{3 \zeta(3)}{2} \epsilon^2  +  O(\epsilon^{5/2}) .
\end{equation}
We note that the anomalous dimensions $\mu_2^{(0)}$ and $\zeta_2^{(0)}$ are different only by sign up to the third-loop order. Their absolute values become different only at the forth loop order. We mention that at the one-loop approximation the anomalous dimensions $\mu_2^{(0)}$ and $\zeta_2^{(0)}$ for noninteracting case coincide with the anomalous dimensions $\zeta_z$ and $\zeta_2$ for the case of Coulomb interaction. Therefore, one can expect that the eigen operator bilinear in $Q$ with the anomalous dimension $\mu_2^{(0)}$ transforms into the Finkel'stein operator in the case of Coulomb interaction.

\section{Conclusions \label{Sec:Conc}}

To summarize, we studied the two-loop renormalization of the Finkel'stein parameter $z_\omega$ which anomalous dimension controls the scaling of the frequency and the specific heat for an interacting disordered electron system. For simplicity, we considered the case of broken time reversal symmetry in order to avoid additional difficulty due to the Cooper channel. Under this assumption we derived the two-loop RG equation for $z_\omega$ (see Eq. \eqref{eq:ad:2l}) the right hand side of which depends on the interaction parameters in the singlet and triplet channels. Our result \eqref{eq:ad:2l} interpolates between the result of Ref. \cite{Baranov1999} for $\gamma_t=0$ and the result of Refs. \cite{Kirkpatrick1989,Kirkpatrick1990} for $\gamma_t\to \infty$. We consider the metal-insulator transition in $d=2+\epsilon$ dimensions in the case of the symmetry class MI(LR), i.e. with broken time reversal and spin rotational symmetries and in the presence of Coulomb interaction. We compared the anomalous dimension of the Finkel'stein operator $S_F$ in the NLSM action with the anomalous dimension of the second moment of the local density of states. Within one-loop approximation both anomalous dimensions coincide in absolute value and are equal to ones in the absence of interaction. However, in the two-loop approximation their absolute values deviate from each other and from anomalous dimensions without interaction. 

Finally, we mention that in order to explore the metal-insulator transition in the presence of spin rotational symmetry results for the two-loop renormalization of the spin susceptibility and conductivity are need. They will be published elsewhere.

\section{Acknowledgements}

We are grateful to N. Chtchelkatchev for his interest and valuable help in the initial stage of the research.
The work was funded in part by Russian Foundation for Basic Research under the grant No. 15-32-20176 
and the Dynasty Foundation. 

\appendix
\section{Evaluation of two-loop integrals\label{App1} }

At first, we combine the two-loop contributions \eqref{eq:2l:1} -- \eqref{eq:2l:3} together. Next we rewrite them in the form $\sum_{m>0} \omega_m Y(i\omega_m)$. We used the following identity: 
$\sum_{m,n>0}\min\{m,n\} y(m,n) = \sum_{m,n>0} m \bigl [y(m,n+m)+y(n+m,m)\bigr ]$. Finally, taking the limit $\omega_m \to 0$, we find
\begin{equation}
\Omega^{(2)} = 4 T N_r \sum_{m>0} \omega_m \delta z_\omega ,
\end{equation}
where
\begin{align}
\delta z_\omega = &  \frac{16 \pi T z_\omega^2}{g}\left (\frac{4}{g}\right )^2 \int_{p,q} \sum_{n>0} \Biggl \{\frac{3\gamma_t+\gamma_s}{4} 
\sum_{j=0}^3 \Bigl [\gamma_j
\mathcal{D}^{(j)}_p(i\omega_n)\mathcal{D}_q^2(0) +\gamma_j \mathcal{D}_p(i\omega_n)\mathcal{D}^{(j)}_p(i\omega_n)\mathcal{D}_q(0) \notag \\
&\hspace{1cm} -\gamma_j
\mathcal{D}(i\omega_n)\mathcal{D}^{(j)}_p(i\omega_n)\mathcal{D}_q(i\omega_n)-2\gamma_j
\mathcal{D}^{(j)}_p(i\omega_n)\mathcal{D}_q(0)\mathcal{D}_{\bm{p}+\bm{q}}(i\omega_n)  \notag \\
&\hspace{1cm} +\gamma_j^2\omega_n \mathcal{D}^{(j)}_p(i\omega_n)\mathcal{D}^2_q(0)\mathcal{D}_{\bm{p}+\bm{q}}(i\omega_n)-\gamma_j
\mathcal{D}_p(i\omega_n) \mathcal{D}^2_q(0)\Bigr ] \notag \\
& \hspace{1cm}  +3\gamma_t^2\Bigl [ - \mathcal{D}_p(i\omega_n)\mathcal{D}^{(j)}_p(i\omega_n) \mathcal{D}_q(0) +
 \mathcal{D}_p(i\omega_n)\mathcal{D}^{(j)}_p(i\omega_n)\mathcal{D}_q(i\omega_n) \notag \\
 & \hspace{1cm} +
\mathcal{D}^{(j)}_p(i\omega_n)\mathcal{D}_q(0)\mathcal{D}_{\bm{p}+\bm{q}}(i\omega_n)+\mathcal{D}_p(i\omega_n)\mathcal{D}_q(0)\mathcal{D}_{\bm{p}+\bm{q}}(i\omega_n)\Bigr
] \Biggr \}\notag \\
& - \frac{8 \gamma_s z_\omega}{g^2}  \int_{p,q}  \mathcal{D}_p(0)\mathcal{D}_q(0)  .
\label{eq:app:1}
\end{align}
Now we set the temperature to zero and will study the dependence of $\delta z_\omega$ on the momentum scale $h$ only. Then we find
\begin{align}
\delta z_\omega &= \frac{16 z_\omega}{g^2} \Biggl \{ \frac{3\gamma_t+\gamma_s}{4} \sum_{j=0}^3 \gamma_j \Bigl [J^0_{020}(\gamma_j) +J^0_{110}(\gamma_j) -
J^0_{101}(\gamma_j)-2J^0_{011}(\gamma_j)+\gamma_j
J^1_{021}(\gamma_j)-J^0_{020}(0)\Bigr ] \notag
\\
& \hspace{1cm} - 3\gamma_t^2\Bigl [
J^0_{110}(\gamma_t) -
 J^0_{101}(\gamma_t) -
J^0_{011}(\gamma_t)- J^0_{011}(0)\Bigr ] -\frac{\gamma_s}{2} J_0 \Biggr \}  .
\label{eq:app:2}
\end{align}
Here we introduced 
\begin{equation}
J^{\delta}_{\nu\mu\eta}(\gamma_j) = \left (\frac{8 T z_\omega}{g}\right )^{1+\delta} \int_{p,q} \int\limits_0^\infty d\omega \,
\omega^\delta
\mathcal{D}^\nu_p(i\omega)\mathcal{D}^{(j)}_p(\omega)\mathcal{D}_q^\mu(0)\mathcal{D}_{\bm{p}+\bm{q}}^\eta(i\omega) 
\end{equation}
and
\begin{equation}
J_0 = \left [ \int_{p} \mathcal{D}_p(0) \right ]^2 .
\end{equation}

\subsection{The integrals $J_0$, $J^0_{020}$ and $J^0_{110}$}

Using the result
\begin{equation}
\int_q \mathcal{D}_q(0) = -\frac{2 \Omega_d h^{\epsilon}
\Gamma(1+\epsilon/2)\Gamma(1-\epsilon/2)}{\epsilon}  ,
\end{equation}
we find
\begin{equation}
J_0 = \frac{4 A_\epsilon h^{2\epsilon}}{\epsilon^2}, \qquad A_\epsilon = \Omega_d^2
\Gamma^2(1-\epsilon/2)\Gamma^2(1+\epsilon/2) .
\label{eq:app:J0} 
\end{equation}

Next, with the help of the results
\begin{equation}
\begin{split}
\int_q \mathcal{D}^2_q(0) & = \Omega_d
h^{\epsilon-2}\Gamma(1+\epsilon/2)\Gamma(1-\epsilon/2), \\
\frac{8 z_\omega}{g} \int\limits_0^\infty d\omega \int_q \mathcal{D}_q^{(j)}(i\omega) & = \frac{4\Omega_d
h^{\epsilon+2}
\Gamma(1+\epsilon/2)\Gamma(1-\epsilon/2)}{(1+\gamma_j)\epsilon(2+\epsilon)} ,
\end{split}
\end{equation}
we obtain
\begin{equation}
J^0_{020}(\gamma_j) = \frac{4A_\epsilon
h^{2\epsilon}}{(1+\gamma_j)\epsilon(2+\epsilon)} .
\label{eq:app:J0_020}
\end{equation}

Using the following relation between integrals
\begin{equation}
\frac{8 z_\omega}{g} \int\limits_0^\infty d\omega \int_q \mathcal{D}_q(i\omega)\mathcal{D}^{(j)}_q(i\omega) = \frac{\ln (1+\gamma_j)}{\gamma_j} \int_q \mathcal{D}_q(0) ,
\end{equation}
we find
\begin{equation}
J^0_{110}(\gamma_j) = \frac{4 A_\epsilon h^{2\epsilon}}{\epsilon^2}
\frac{\ln (1+\gamma_j)}{\gamma_j}  .
\label{eq:app:J0_110}
\end{equation}

\subsection{The integral $J^0_{101}$}

Next we consider the integral $J^0_{101}$. With the help of the Feynman trick (see, e.g. Ref. \cite{Amit-book}), we write
\begin{equation}
J^0_{101}(\gamma_j) = -\frac{A_\epsilon \Gamma(1-\epsilon)
h^{2\epsilon}}{\epsilon \Gamma^2(1-\epsilon/2)} T_{01}(\gamma_j) ,
\end{equation}
where (see Eq. (A26) of Ref. \cite{Burmistrov2002})
\begin{equation}
T_{01}(\gamma_j)= \left ( \prod_{k=1}^3 \int\limits_0^1 dx_k \right ) \delta\left (\sum_{k=1}^3 x_k -1\right )  \frac{x^{-1-\epsilon/2}_3
x_{12}^{-1-\epsilon/2}}{x_{1}+(1+\gamma_j) x_2 + x_3} .
\end{equation}
Performing the change of variables from $x_1, x_2$ and $x_3$ to $s$ and $u$:
\begin{equation}
x_1 = \frac{1-u}{s+1}, \quad x_2 = \frac{u}{s+1}, \quad x_3 = \frac{s}{s+1}, \quad 0\leqslant u \leqslant 1, \quad 0\leqslant s < \infty ,
\label{eq:app:ch}
\end{equation}
and integrating over $s$, we find
\begin{equation}
T_{01}(\gamma_j)=\frac{2\Gamma^2(1-\epsilon/2)}{\gamma_j \epsilon
\Gamma(1-\epsilon)} \int\limits_{1+\gamma_j}^{1}
\frac{du}{u^{1+\epsilon/2}}\, {}_{2}F_{1}(-\epsilon/2,-\epsilon,1-\epsilon,1-u) .
\end{equation}
Here ${}_{2}F_{1}(\alpha,\beta,\gamma,z)$ denotes the hypergeometric function. Rewriting the integral as\begin{align}
T_{01}(\gamma_j)= & -\frac{1}{\gamma_j} \int\limits_{1+\gamma_j}^1 du \Biggl [
\frac{2\Gamma(1-\epsilon/2)\Gamma(1+\epsilon/2)}{-\epsilon}u^{-1-\epsilon/2}
(1-u)^{\epsilon}  \notag 
\\
& \hspace{1.8cm} -
\frac{4\Gamma^2(1-\epsilon/2)}{(2+\epsilon)\Gamma(1-\epsilon)}
{}_{2}F_{1}(1-\epsilon/2,1,2+\epsilon/2,u) \Biggr ] ,
\end{align}
and using that ${}_{2}F_{1}(1-\epsilon/2,1,2+\epsilon/2,u)\to - \ln(1-u)/u$ in the limit $\epsilon \to 0$, we find
\begin{equation}
T_{01}(\gamma_j)= -\frac{\Gamma^2(1-\epsilon/2)}{\gamma_j \Gamma(1-\epsilon)}\left [ \frac{2 \ln (1+\gamma_j)}{\epsilon} -\frac{1}{2}\ln^2(1+\gamma_j)\right ] .
\end{equation}
Hence, we obtain
\begin{equation}
J^0_{101}(\gamma_j) = \frac{A_\epsilon h^{2\epsilon}}{\gamma_j} \left
[\frac{2\ln (1+\gamma_j)}{\epsilon^2} -\frac{\ln^2(1+\gamma_j)}{2\epsilon} \right ] .
\label{eq:app:J0_101}
\end{equation}

\subsection{The integral $J^0_{011}$}

Using the Feynman trick (see, e.g. Ref. \cite{Amit-book}), we write
\begin{equation}
J^0_{011}(\gamma_j)=-\frac{A_\epsilon \Gamma(1-\epsilon)
h^{2\epsilon}}{\epsilon \Gamma^2(1-\epsilon/2)} S_0(\gamma_j) ,
\end{equation}
where (see Eq.(A23) of Ref. \cite{Burmistrov2002})
\begin{equation}
S_0(a) = \left ( \prod_{k=1}^3 \int\limits_0^1 dx_k \right ) \delta\left (\sum_{k=1}^3 x_k -1\right )   \frac{(x_1x_2+x_2 x_3+x_3 x_1)^{-1-\epsilon/2}}{(1+\gamma_j) x_{1}+x_3}.
\end{equation}
Changing variables from $x_1, x_2, x_3$ to $s$ and $u$ (see Eq. \eqref{eq:app:ch}) and evaluating integral over $s$, we obtain
\begin{equation}
S_{0}(\gamma_j) = -\frac{2}{\epsilon}\int\limits_0^1 \frac{du
[u(1-u)]^{-\epsilon/2}}{(1+\gamma_j)u+1-u} {}_2 F_1(1,-\epsilon,1-\epsilon/2;
1-u(1-u)) .
\end{equation}
Rewriting the integral as
\begin{eqnarray}
S_0(\gamma_j) = \frac{2}{\epsilon} \int\limits_0^1 \frac{du
[u(1-u)]^{-\epsilon/2}}{(1+\gamma_j) u+1-u} {}_2 F_1(1,-\epsilon,1-\epsilon/2;
u(1-u))-\frac{4}{\epsilon} \int\limits_0^1 \frac{du
[u(1-u)]^{\epsilon/2}}{(1+\gamma_j) u+1-u} ,
\end{eqnarray}
and using that ${}_2 F_1(1,-\epsilon,1-\epsilon/2; u(1-u)) \to 1 +\epsilon \ln
[1-u(1-u)]$ in the limit $\epsilon\to 0$, we find the following result
\begin{equation}
S_0(\gamma_j) = -\frac{2}{\epsilon}\frac{\ln (1+\gamma_j)}{\gamma_j} - \frac{2}{\gamma_j} \left
[ \liq(-\gamma_j) + \frac{1}{4}\ln^2 (1+\gamma_j)\right ] .
\end{equation}
Here we have used the identities 
\begin{equation}
\begin{split}
\liq(a) -\liq(1)+\ln a\ln(1-a)=-\liq(1-a) , \qquad 0\leqslant a\leqslant 1 ,\\
-\liq(1/a) +\liq(1)+\ln (a-1)\ln a-\frac{1}{2}\ln^2 a=-\liq(1-a), \qquad a > 0 .
\end{split}
\end{equation}
Hence we obtain
\begin{equation}
J^0_{011}(\gamma_j) = \frac{A_\epsilon h^{2\epsilon}}{\gamma_j}\Biggl [
\frac{2\ln (1+\gamma_j)}{\epsilon^2} +\frac{2}{\epsilon} \left
[\liq(-\gamma_j)+\frac{1}{4}\ln^2 (1+\gamma_j)\right ]
 \Biggr
] .
\label{eq:app:J0_011}
\end{equation}

\subsection{The integral $J^1_{021}$}

As above we use the Feynman trick (see, e.g. Ref. \cite{Amit-book}) to write
\begin{equation}
J^1_{021}(a)=-\frac{A_\epsilon \Gamma(1-\epsilon)
h^{2\epsilon}}{\epsilon \Gamma^2(1-\epsilon/2)} S^{1}_{2}(\gamma_j) ,
\end{equation}
where
\begin{equation}
S^1_2(\gamma_j)=\left ( \prod_{k=1}^3 \int\limits_0^1 dx_k \right ) \delta\left (\sum_{k=1}^3 x_k -1\right )   (x_1x_2+x_2x_3+x_3x_1)^{-1-\epsilon/2} \frac{x_2}{ ((1+\gamma_j)
x_{1}+x_3)^2} .
\end{equation}
Changing variables from $x_1, x_2, x_3$ to $s$ and $u$ (see Eq. \eqref{eq:app:ch}) and evaluating integral over $s$, we obtain
\begin{equation}
S^1_2(\gamma_j) = \frac{4}{\epsilon(2+\epsilon)} \int\limits_0^1 \frac{du\,
[u(1-u)]^{1-\epsilon/2}}{((1+\gamma_j)u+1-u)^2}
{}_{2}F_{1}(2,-\epsilon,1-\epsilon/2,1-u(1-u)) .
\end{equation}
Rewriting the integral as
\begin{gather}
S^1_2(\gamma_j)=
\int\limits_0^1 \frac{du\, [u(1-u)]^{1-\epsilon/2}}{((1+\gamma_j) u+1-u)^2}\Biggl  [
-\frac{2}{2+\epsilon}{}_{2}F_{1}(1,-\epsilon,-\epsilon/2,1-u(1-u)) \notag \\
+\frac{2}{\epsilon} {}_{2}F_{1}(1,-\epsilon,1-\epsilon/2,1-u(1-u))
\Biggr ] ,
\end{gather}
and using that ${}_{2}F_{1}(1,-\epsilon,-\epsilon/2,1-u(1-u)) \to
[2-u(1-u)]/[u(1-u)]$  in the limit $\epsilon\to 0$, we obtain the following result
\begin{gather}
S_2^1(\gamma_j)=-\frac{2}{\epsilon} \frac{2\gamma_j- (2+\gamma_j)\ln (1+\gamma_j)}{\gamma_j^3}
-\frac{2}{1+\gamma_j} + \frac{2(2+\gamma_j)\ln(1+\gamma_j)}{\gamma_j^3} \notag \\
+\frac{2(2+\gamma_j)}{\gamma_j^3}
\left [\liq(-\gamma_j)+\frac{1}{4}\ln^2 (1+\gamma_j) \right ] .
\end{gather}
Hence, we find
\begin{gather}
J^1_{021}(\gamma_j) = A_\epsilon h^{2\epsilon}\Biggl \{
 \frac{4\gamma_j-2(2+\gamma_j)\ln (1+\gamma_j)}{\gamma_j^3\epsilon^2} +\frac{2}{(1+\gamma_j)\epsilon}
- \frac{2(2+\gamma_j)\ln (1+\gamma_j)}{\gamma_j^3\epsilon}\notag \\ -\frac{2(2+\gamma_j)}{\gamma_j^3\epsilon}
\left [\liq(-\gamma_j)+\frac{1}{4}\ln^2 (1+\gamma_j)\right ]
  \Biggr  \} .
  \label{eq:app:J1_021}
\end{gather}

Finally, substituting the results for the integrals into Eq. \eqref{eq:app:2}
 we obtain the result \eqref{eq:Z:pert:2l}.

\vspace{1cm}

\bibliographystyle{elsarticle-num-names}
\bibliography{2d-biblio-v1.bib}

\end{document}